\documentclass[aps,twocolumn,prl,reprint,superscriptaddress,showpacs]{revtex4}
\usepackage{amsmath}
\usepackage{graphicx}
\usepackage{bm}
\usepackage{amssymb}
\usepackage{hyperref}
\begin{document}
\title{$\pi$ Spin Berry Phase in a Quantum-Spin-Hall-Insulator-Based Interferometer:
Evidence for the Helical Spin Texture of the Edge States}
\author{Wei Chen\footnote{These authors contributed equally to this work.}}
\email{weichenphy@nuaa.edu.cn}
\affiliation{College of Science, Nanjing University of Aeronautics and Astronautics, Nanjing 210016, China}
\author{Wei-Yin Deng$^*$}
\affiliation{National Laboratory of Solid State Microstructures, Department of Physics, and Collaborative Innovation Center of Advanced Microstructures, Nanjing University, Nanjing 210093, China}
\author{Jing-Min Hou}
\affiliation{Department of Physics, Southeast University, Nanjing 211189, China}
\author{D. N. Shi}
\affiliation{College of Science, Nanjing University of Aeronautics and Astronautics, Nanjing 210016, China}
\author{L. Sheng}
\affiliation{National Laboratory of Solid State Microstructures, Department of Physics, and Collaborative Innovation Center of Advanced Microstructures, Nanjing University, Nanjing 210093, China}
\author{D. Y. Xing}
\affiliation{National Laboratory of Solid State Microstructures, Department of Physics, and Collaborative Innovation Center of Advanced Microstructures, Nanjing University, Nanjing 210093, China}
\begin{abstract}
Quantum spin Hall insulator is characterized by the helical edge states, with the spin polarization of electron being locked to its direction of motion. Although the edge-state conduction has been observed, unambiguous evidence of the helical spin texture is still lacking. Here, we investigate the coherent edge-state transport in an interference loop pinched by two point contacts. Due to the helical character, the forward inter-edge scattering enforces a $\pi$ spin rotation. Two successive processes can only produce a nontrivial $2\pi$ or trivial $0$ spin rotation, which can be controlled by the Rashba spin-orbit coupling. The nontrivial spin rotation results in a geometric $\pi$ Berry phase, which can be detected by a $\pi$ phase shift of the conductance oscillation relative to the trivial case. Our results provide a smoking gun evidence for the helical spin texture of the edge states. Moreover, it also provides the opportunity to all-electrically explore the trajectory-dependent spin Berry phase in condensed matter.
\end{abstract}
\pacs{73.20.-r, 03.65.Vf, 85.75.-d, 73.23.-b}
\maketitle

Quantum spin Hall insulator (QSHI) is a topologically nontrivial phase of electronic matter \cite{Kane,Zhang,Bernevig,Molenkamp,Roth,Kane2}, which is characterized by the gapped bulk states and helical edge states \cite{Wu}, i.e., the spin-up (spin-down) electrons propagate clockwise (counterclockwise) along the sample edge. Besides the physical significance, the helical edge states may have important applications in the spintronic device \cite{Kim,Richter,Dolcini} and quantum information processing \cite{Loss,Chen}. Although the edge-state conduction \cite{Molenkamp,Roth} and the spin polarization of the edge current \cite{Brune} have been confirmed, a direct evidence of the helical spin texture of the edge states remains a challenge, due to the lack of experimental probe that can measure the key property for a one dimensional edge mode at a buried interface at mK temperatures \cite{Hasan}. In this work, the spin-resolved coherent edge transport in a QSHI-based interferometer is investigated. Interestingly, a $\pi$ spin Berry phase can be induced by the inter-edge scattering for the helical edge states,
which can serve as the smoking gun evidence of their existence. At the same time, our scheme also provides the opportunity to observe the trajectory-dependent spin evolution and the corresponding spin Berry phase \cite{BF1}, which is also of great importance in condensed matter physics \cite{BF1,BF2,BF3,BF4,BF5,BF6,BF7,BF8}. The existing evidences of spin Berry phase usually come from the anomalous Hall effect \cite{BF2,BF3}, Shubnikov-de Haas effect \cite{BF1,BF4,BF5,BF6,BF7,BF8}, and weak anti-localization effect \cite{Ando1,Ando2}. In these experiments, a clear picture of spin evolution is hard to extract and a magnetic field is always needed. Remarkably, by utilizing the helical edge states of the QSHI, the detection of spin Berry phase can be achieved all-electrically.

\begin{figure}
\centering
\includegraphics[width=0.48\textwidth]{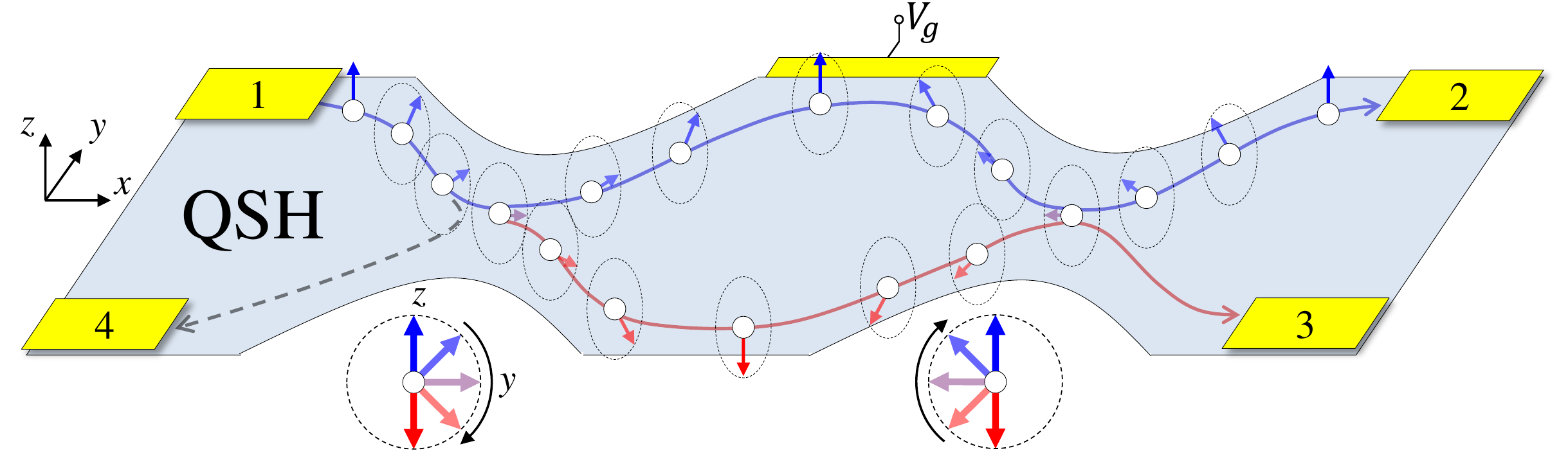}
\caption{(Color). The interference loop for detecting the $\pi$ spin Berry phase in the QSHI, in which there are four terminals (yellow bars) and two point contacts (PCs). The trajectory-dependent spin evolution in the $y$-$z$ plane along two coherent paths is sketched. A side gate $V_g$ is fabricated at the upper edge, to modulate the interference pattern of the conductance.
 } \label{fig1}
\end{figure}

The proposed setup fabricated on the QSHI is shown in Fig. \ref{fig1}. Due to the spin-momentum locking, the electrons moving to the right along the upper and lower edge channels must have opposite spin directions. Spin-up electrons starting from terminal 1 first go forward along the upper edge channel, then get scattered at two PCs (narrow regions), and finally may reach  terminals 2, 3, or 4. The effect due to direct connection between terminals 1 and 4 through the edge channels on the left side can be eliminated by isolating the region of interest with additional contacts \cite{Edge}, as was done in the quantum Hall system \cite{Henny,Bocquillon}. By imposing a perpendicular electric field $E_z$ around each PC, the Rashba spin-orbit coupling (SOC) can be generated \cite{Rothe,Japaridze,Ojanen}, which induces the forward inter-edge scattering. Due to the helical nature of the edge states, such a scattering process can occur only when the electronic spin rotates by $\pi$ simultaneously. If we focus on the conduction from terminal 1 to 2, the dominant coherent paths between the two PCs are the upper spin-up edge channel and the lower spin-down edge channel. For the latter, the electron will undergo two successive forward inter-edge scattering processes with spin reversal, and so give rise to either $2\pi$ or $0$ spin rotation, depending on whether the two spin rotations are in the same or opposite direction, which can be tuned by the sign of the Rashba coefficients. Such a $2\pi$ spin rotation,  shown by two dashed line circles below the setup in Fig. \ref{fig1},  is highly nontrivial and can induce an additional $\pi$ Berry phase to the wave function \cite{Sakurai}. Accordingly, for the coherent transport between terminal 1 and 2, the amplitudes through the lower path with nontrivial $2\pi$ and trivial $0$ spin rotations exhibit a $\pi$ phase difference. This can be revealed by the $\pi$ phase shift of the interference pattern of conductance $G_{12}$, as the Rashba coefficient at one PC changes the sign. Since the $2\pi$ and $0$ spin rotations are enforced by the helical nature of the edge states, such a $\pi$ spin Berry phase provides an unambiguous evidence of the helical spin texture of the edge states.

In the following, we first adopt an effective model to present a clear picture of the trajectory-dependent spin Berry phase, and then perform numerical calculation to give rigorous results. The QSHI phase is stabilized by the intrinsic SOC, which induces helical edge states according to the bulk-edge correspondence \cite{Bernevig, Kane2}. Within the energy gap of the bulk material, the system can be described by the effective one dimensional Hamiltonian as
\begin{equation}\label{H0}
H_0=-iv_F\partial_x\sigma_z\tau_z-i\alpha(x)\partial_x\sigma_y+\beta(x)\tau_x+\varepsilon_0,
\end{equation}
where the first term describes the helical edge states \cite{Wu} with $v_F$ being the Fermi velocity, the second term is the Rashba SOC term \cite{Japaridze}, the third term introduces an effective inter-edge coupling due to the finite size effect \cite{Zhou}, and the constant $\varepsilon_0=7.5$ meV is the energy of the Dirac point. The Pauli matrices $\sigma_{x,y,z}$ and $\tau_{x,y,z}$  operate  in the spin and edge spaces, respectively. Both the Rashba SOC coefficient $\alpha(x)$ and the inter-edge coupling strength $\beta(x)$ are nonuniform and take finite values only around the PCs. The Rashba coefficient $\alpha(x)$ is assumed smooth enough, so that the commute relation $[\partial_x,\alpha(x)]=0$ holds, which ensures the Hermitian of the Hamiltonian (\ref{H0}) \cite{note}. In addition, the Rashba coefficient satisfies $\alpha\propto eE_z$, so that the sign of $\alpha$ can be changed by inverting the electric field.

The Rashba SOC is equivalent to a nonuniform Zeeman field, which shifts the spin direction of the helical edge states. It is convenient to solve the Schr\"{o}dinger equation $H_0\psi=E\psi$ under the rotated basis $\Psi=U(\theta)\psi$, where the unitary operator $U(\theta)=\exp(-i\sigma_x\tau_z\theta/2)$ introduces opposite spin rotations $R_x(\pm\theta)$ about the $x$ axis to the upper and lower edges, with the angle being defined by $\theta=\tan^{-1}(\alpha/v_F)$. Under the rotated basis, the Hamiltonian $\mathcal{H}=U(H_0-\varepsilon_0)U^\dag$ writes
\begin{equation}\label{H}
\mathcal{H}=-i\mathcal{V}_F\partial_x\sigma_z\tau_z+\beta(\cos\theta\tau_x+\sin\theta\sigma_x\tau_y),
\end{equation}
where the energy is measured from $\varepsilon_0$ and the Fermi velocity is renormalized to $\mathcal{V}_F=\sqrt{v_F^2+\alpha^2}$. The Rashba SOC induces the spin flipped inter-edge scattering with a strength of $\beta\sin\theta$, where $\sin\theta$ is the overlap between the spin states at different edges moving in the same direction.

In order to investigate the coherent transport through the whole interferometer, we first solve the scattering problem at one PC, and then combine the scattering amplitudes of two PCs. Note that for a right-moving wave packet with a wave vector satisfying $k\gg\beta/\mathcal{V}_F$, its coupling to the left-moving states with an energy difference of $2\mathcal{V}_Fk$ can be safely neglected. The first order of the inter-edge backscattering amplitude is $\beta\cos\theta/(2\mathcal{V}_Fk)$ according to the perturbation theory. In this regime, the counter-moving electron states are decoupled, and the right-moving Hamiltonian reduces to
\begin{equation}\label{HR}
\mathcal{H}_\rightarrow=-i\mathcal{V}_F\partial_x+\beta\sin\theta\tau_y,
\end{equation}
which describes the coupling between the degenerate right-moving states at opposite edges. Due to the helical nature of the edge states, the spin index is redundant.

Let the center of the left PC be located at $x=0$, and the inter-edge coupling $\beta(x)$ takes non-zero value in the region $(-\frac{L_{PC}}{2},\frac{L_{PC}}{2})$, with $L_{PC}$ being the length of the PC. We also assume that $\alpha(x)$ varies much more slowly than $\beta(x)$, so that the change of $\alpha(x)$ around the PC can be neglected, and the rotation angle $\theta(x)=\theta_0$ is a constant.

For an electron incident from $x=-\infty$ with an initial spin state of $\psi_i=(a_1|\uparrow\rangle, a_2|\downarrow\rangle)^T$, it first undergos a spin rotation due to the Rashba SOC, and the spin state becomes $\psi_1=U(-\theta_0)\psi_i$ as it reaches the left side of the PC. Then the electron gets forward scattered at the PC region, and the state changes to $\psi_2=S\psi_1$, where the scattering matrix $S$ is solved by the Hamiltonian (\ref{HR}) \cite{SM}. Finally, after the electron transmits the PC, the regression of the Rashba SOC pulls the spin polarization direction back to the $z$ axis, and the final spin state writes $\psi_f=U(\theta_0)\psi_2$. The whole process can be written in a compact form as $\psi_f=\mathcal{S}\psi_i$, with the matrix $\mathcal{S}=U(\theta_0)SU(-\theta_0)$ being expressed as \cite{SM}
\begin{equation}\label{S}
\mathcal{S}(\pi)=\left(
                             \begin{array}{cc}
                               \cos\phi & -i\sin|\phi| R_x(\pi) \\
                               -i\sin|\phi| R_x(-\pi) & \cos\phi \\
                             \end{array}
                           \right),
\end{equation}
where the phase angle is defined by
\begin{equation}\label{phi}
\phi=\frac{\sin\theta_0}{\mathcal{V}_F}\int_{-\frac{L_{PC}}{2}}^\frac{L_{PC}}{2}\beta(x)dx.
\end{equation}
The matrix $\mathcal{S}$ provides a full description of the forward scattering processes at one PC. Its diagonal and off-diagonal elements describe the intra-edge and inter-edge scattering, respectively.
The helical character of the edge states is revealed by the $\pi$ spin rotation $R_x(\pm\pi)$, which means that the forward inter-edge scattering occurs only when the spin is flipped by an angle of $\pi$. This property is unique for the helical states, while for the normal states, the spin conserved forward inter-edge scattering is also allowed. For the intra-edge scattering, the spin first deviates from the $z$ direction, and then gets compensated, finally leaving the spin state unchanged. If the Rashba coefficient changes its sign as $\alpha\rightarrow -\alpha$, the precession direction of the spin is inverted as well. As a result, the corresponding scattering matrix changes to $\mathcal{S}(-\pi)$.

For a single PC, the inter-edge scattering induces a $\pi$ spin rotation, so that in an interference loop composed of two PCs, two such processes can only produce $2\pi$ or $0$ spin rotation. To observe this effect, a side gate $V_g$ is deposited on the upper edge in the middle region as shown in Fig. \ref{fig1}. It introduces a phase difference of $\delta=eV_gL_G/v_F$ between the upper and lower edge channels, with $L_G$ being the length of the gating region. The effect of the side gate can be expressed by a transfer matrix as
$\mathcal{T}=\left(
\begin{array}{cc}
e^{i\frac{\delta}{2}} & 0 \\
0 & e^{-i\frac{\delta}{2}} \\
\end{array}
\right)
$. For the nontrivial case, the Rashba coefficients take opposite signs at the left and right PCs and the corresponding matrices $\mathcal{S}_L(\pi)$ and $\mathcal{S}_R(-\pi)$ are defined by Eq. (\ref{S}). In this case, the scattering matrix $\mathcal{M}$ for the whole interferometer is obtained as $\mathcal{M}=\mathcal{S}_R(-\pi)\mathcal{T}\mathcal{S}_L(\pi)$, and a direct calculation yields
\begin{equation}
\begin{split}
\mathcal{M}&=\left(
                             \begin{array}{cc}
                              \lambda & -i\chi \\
                               -i\chi^* & \lambda^* \\
                             \end{array}
                           \right),\\
\lambda&=e^{i\frac{\delta}{2}}\cos\phi_L\cos\phi_R -e^{-i\frac{\delta}{2}}\sin|\phi_L|\sin|\phi_R| R_x(-2\pi),\\
\chi&=e^{i\frac{\delta}{2}}\sin|\phi_L|\cos\phi_R R_x(\pi)+e^{-i\frac{\delta}{2}}\cos\phi_L\sin|\phi_R| R_x(-\pi),
\end{split}
\end{equation}
where the phase angles $\phi_{L,R}$ defined in Eq. (\ref{phi}) correspond to the left and right PCs, respectively. One can see that the second term of the diagonal element $\lambda$ contains a $2\pi$ rotation, which comes from two successive forward inter-edge scattering. Such spin rotation contributes a Berry phase of $R_x(-2\pi)=e^{i\pi}$, which can be observed through the coherent oscillation of the conductance.

Here we focus on the electron conduction between terminal 1 and 2. According to the Landauer-B\"{u}ttiker formula, the reduced conductance (in units of $G_0=e^2/h$) equals $G_{12}/G_0=|\lambda|^2$. Inserting the expression of $\lambda$ yields
\begin{equation}
\begin{split}
G_{12}/G_0&=\cos^2\phi_L\cos^2\phi_R+\sin^2\phi_L\sin^2\phi_R\\
&-\frac{1}{2}\sin2|\phi_L|\sin2|\phi_R|\cos(\delta-\pi).
\end{split}
\end{equation}
The conductance exhibits an oscillating behavior as the phase $\delta$ varies, which can be tuned by the gate voltage $V_g$. More importantly, a $\pi$ phase factor exists in the expression of the conductance, which stems from the spin rotation $R_x(-2\pi)$. Such $\pi$ spin Berry phase can be observed by taking the conductance pattern in the trivial case as reference, where two Rashba coefficients at two PCs take the same sign. In this case, the scattering matrix is $\tilde{\mathcal{M}}=\mathcal{S}_R(\pi)\mathcal{T}\mathcal{S}_L(\pi)$, and the conductance is $\tilde{G}_{12}/G_0=\cos^2\phi_L\cos^2\phi_R+\sin^2\phi_L\sin^2\phi_R-\frac{1}{2}\sin2|\phi_L|\sin2|\phi_R|\cos\delta$. It is clear that the interference patterns of the conductance for the nontrivial and trivial cases possess a $\pi$ phase shift.

\begin{figure}
\centering
\includegraphics[width=0.48\textwidth] {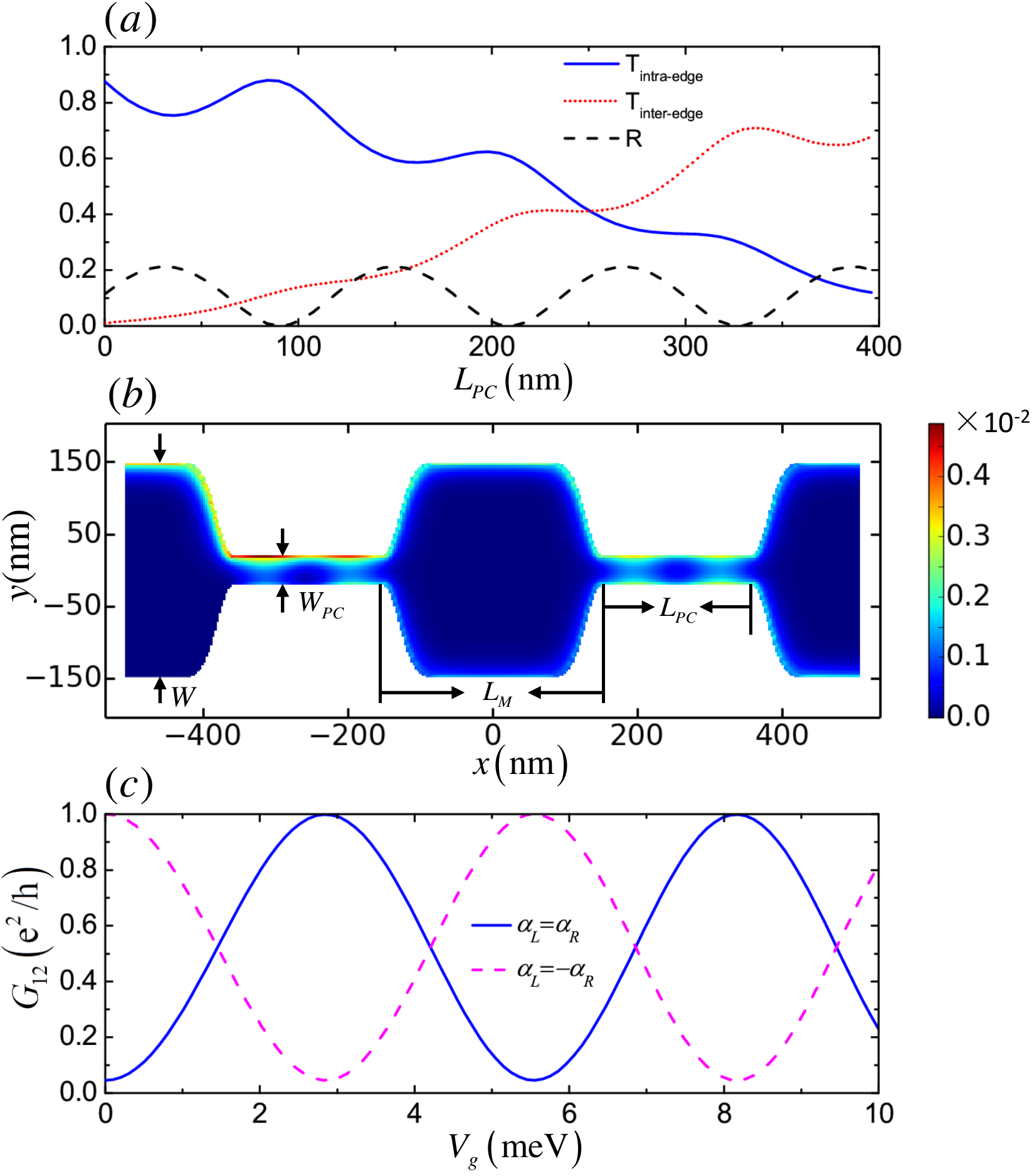}
\caption{(Color). The geometric parameters (see (b)) for one PC are set as $W=300{\rm{nm}}$, $W_{PC}=40{\rm{nm}}$. For the interferometer composed of two PCs, the length of the middle region is $L_M=300\rm{nm}$. The region with finite Rashba SOC covers the PC length $L_{PC}$, and extends by a distance of $\delta L=25{\rm{nm}}$ on both sides. The side gate $V_g$ is applied on the upper edge of the middle region within $3W/8 \le y \le W/2$. (a) The scattering probabilities at one PC as a function of $L_{PC}$ for an incident energy of $E_0=0.0{\rm{meV}}$. The Rashba coefficient is chosen as $\alpha=160{\rm{nm\ meV}}$. (b) The density distribution of the edge transport in the whole interferometer with $L_{PC}=210{\rm{nm}}$ and $V_g=1.4{\rm{meV}}$. (c) The conductance $G_{12}$ as a function of $V_g$. The blue solid line and pink dashed line correspond to the same sign ($\alpha_L=\alpha_R=160$ nm meV) and opposite sign ($\alpha_L=-\alpha_R=160$ nm meV) of the Rashba coefficients at two PCs, respectively. When the sign of the right Rashba coefficient changes, the conductance oscillation exhibits a $\pi$ phase shift.} \label{fig2}
\end{figure}

From the above analysis, we have seen that a nontrivial spin Berry phase during the inter-edge scattering can be observed by the conductance oscillation. In the analytic calculation, several assumptions have been made, such as the slowly varying of $\alpha(x)$ and $\beta(x)$. Next we perform numerical calculation to give rigorous results, based on the lattice model of the HgTe/CdTe quantum wells \cite{SM}, by using the Kwant package \cite{Groth}. It turns out that the main results still hold, independent on these assumptions.

The shape and geometric parameters of the whole system are given in Fig. \ref{fig2}(b). To observe the $\pi$ phase shift of the conduction $G_{12}$, it is advantageous to eliminate the backscattering, so that the conduction between terminal 1 and 2 are purely contributed by two forward scattering paths as shown in Fig. \ref{fig1}. Since the intra-edge backscattering is prevented by the time reversal symmetry \cite{Kane2}, the only possibility of backscattering is the inter-edge backscattering. In Fig. \ref{fig2}(a), the probabilities of the forward scattering $\text{T}_{\text{intra-edge}}, \text{T}_{\text{inter-edge}}$ and the inter-edge backscattering $\text{R}$ around a single PC as a function of the PC length $L_{PC}$ are plotted, where R exhibits an oscillating behavior as $L_{PC}$ increases. It can be verified that other parameters such as the Rashba coefficient $\alpha(x)$ have little effect on the backscattering \cite{SM}. As $L_{PC}$ is chosen properly, $\text{R}=0$ can be achieved. Here, $L_{PC}=210$ nm is chosen, such that the scale of the whole interferometer composed of two PCs is still within the coherent length of the edge states \cite{Molenkamp, Roth}. In Fig. \ref{fig2}(b), the density distribution of the edge transport is given. The conductance $G_{12}$ as a function of the voltage $V_g$ of the side gate is shown in Fig. \ref{fig2}(c). Remarkably, as the Rashba coefficient for the right PC changes its sign, a $\pi$ phase shift of the conductance pattern occurs, which provides an unambiguous evidence of the helical edge states. Such $\pi$ phase shift can be observed as long as the backscattering is absent. To obtain high resolution of the interference pattern, $\text{T}_{\text{intra-edge}}\simeq\text{T}_{\text{inter-edge}}$ is favorable. In Fig. \ref{fig2}(c), the same geometric parameters $L_{PC}, W_{PC}$ and the same Rashba SOC strength $|\alpha_L|=|\alpha_R|$ for two PCs are adopted. However, this is not essential to observe the $\pi$ phase shift. In the Supplemental Material \cite{SM}, we show that the spin Berry phase can be observed in a wide range of these parameters, which facilitates the experimental realization.

\begin{figure}
\centering
\includegraphics[width=0.48\textwidth]{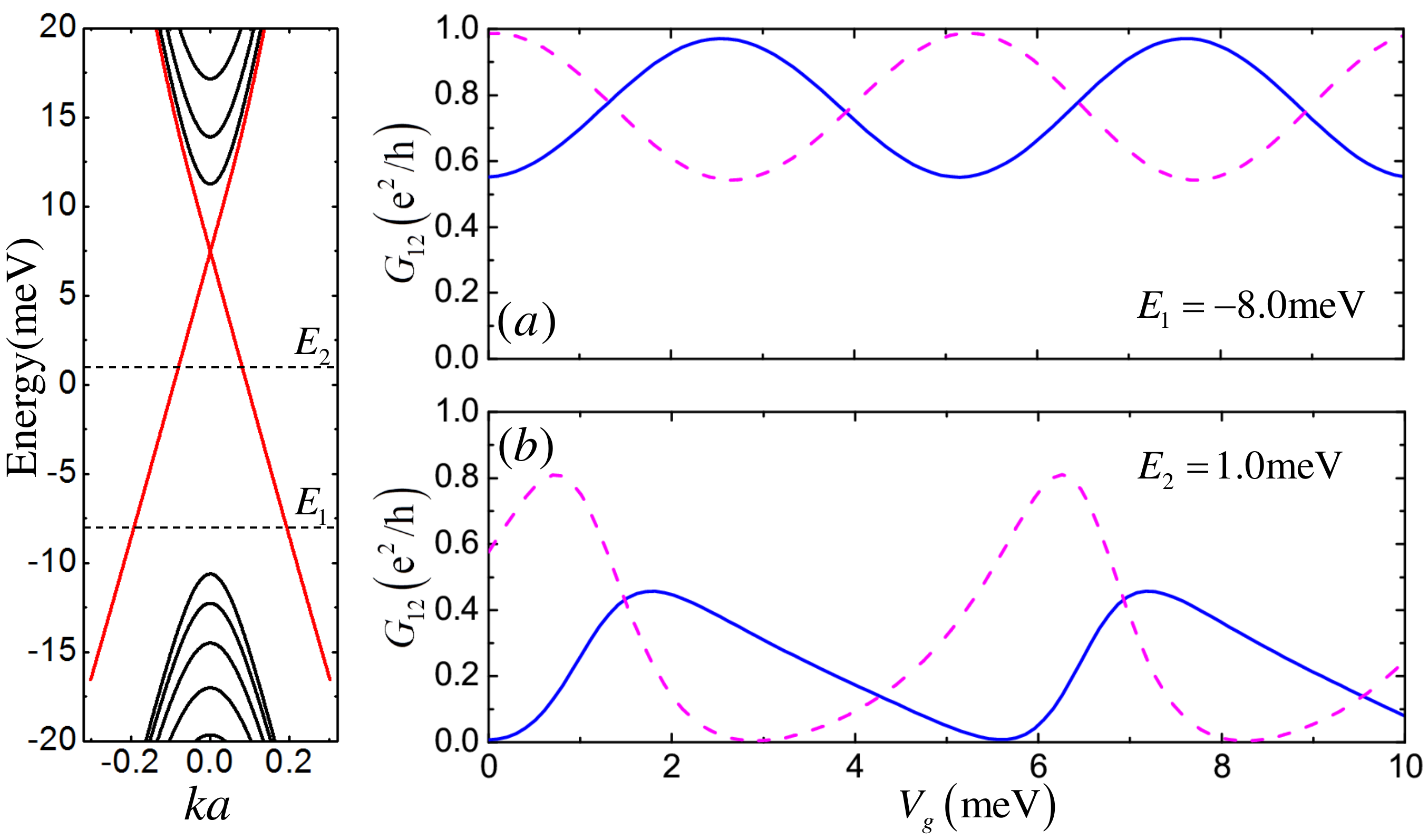}
\caption{(Color). Left: The band structure of the HgTe/CdTe quantum well with a width of $W=300{\rm{nm}}$. The red and black lines correspond to the edge and bulk states, respectively. Right: The conductance $G_{12}$ as a function of $V_g$ for an incident energy of (a) $E_1=-8.0 \rm{meV}$ and (b) $E_2=1.0\rm{meV}$. All other parameters and line settings are the same as that in Fig. \ref{fig2}(c).} \label{fig3}
\end{figure}

Besides the geometric parameters, there are two crucial energy levels for the edge transport. To suppress the backscattering, the incident energy of electron should stay far away from both the Dirac point and the band edge of the bulk states. Otherwise, the backscattering can not be neglected, which will lead to complex interference between numberless coherent paths and smear the harmonic interference pattern and the $\pi$ phase shift. In Fig. \ref{fig3}, we investigate the effect of the incident energy on the spin Berry phase. The incident energy in Fig. \ref{fig2}(c) is $E_0=0.0$ meV, which lies well in the middle of the Dirac point and the edge of the bulk valence band. In Fig. \ref{fig3}(a), the incident energy is $E_1=-8.0$ meV, and the corresponding interference pattern exhibits lower resolution, due to the inequality of $\text{T}_{\text{intra-edge}}$ and $\text{T}_{\text{inter-edge}}$. However, the $\pi$ phase shift is still explicit, since the backscattering remains suppressed. For the incident energy $E_2=1.0$ meV in Fig. \ref{fig3}(b), the harmonic oscillation of the conductance and the $\pi$ phase shift are absent. This stems from the intervention of the inter-edge backscattering as the energy approaches the Dirac point.

It is worthwhile to discuss the experimental realization of our proposal. The whole setup based on the HgTe/CdTe quantum wells can be fabricated and structured by the mature technologies \cite{Molenkamp}. The side gate can be realized by the gate insulator stack, and the Rashba SOC can be induced by the gate insulators around two PCs \cite{Molenkamp}. The Rashba coefficient with a magnitude of $\alpha=160{\rm{nm\ meV}}$ in our calculation is achievable as well \cite{Rothe,Ojanen}. Although our calculation is based on the model of HgTe/CdTe quantum wells, the general physics should hold for any QSHI systems, such as InAs/GaSb quantum wells \cite{Liu,Du}. In order to observe the $\pi$ spin Berry phase, three points should be addressed. (i) The key point is to suppress the backscattering at the PCs, which can be achieved by choosing proper geometric parameters $W_{PC}$ and $L_{PC}$ of the PCs. It is shown that a wide range of these parameters are available \cite{SM}. (ii) To verify zero backscattering, it is convenient to measure the conductance $G_{14}$ between the terminal 1 and 4 in Fig. \ref{fig1}. By setting a proper incident energy of the electron, $G_{14}=0$ should hold for any $V_g$, so that the accidental case due to the interference effect for certain $V_g$ can be excluded. (iii) As long as zero backscattering gets confirmed, the spin Berry phase can be extracted by the $\pi$ phase shift between two conductance patterns $G_{12}$ corresponding to opposite Rashba coefficients $\alpha_R$ and $-\alpha_R$.

There exist several proposals on the detection of the helical spin texture. In Refs. \cite{Edge,Schmidt,Romeo}, the current-current correlation was proposed to be measured; in Refs. \cite{Das,Soori}, either spin-polarized scanning-tunneling probe or magnetic field was prerequisite; in Refs. \cite{Hou,Adroguer} the electron-electron interaction strength should be tuned. Compared with these proposals, our work provides the most direct evidence of the helical spin texture, which can be detected all-electrically through the conductance patterns. Our work also provides a promising scheme to explore the trajectory-dependent spin Berry phase in condensed matter.

\begin{acknowledgments}
We would like to thank Ming Gong and Tao Zhou for helpful discussions. This work was supported by the National Natural Science Foundation of China under Grant No. 11504171 (W.C.), No. 11274061 (J.M.H.), No. 11374159 (D.N.S.) and No. 11225420 (L.S.). W.C. also acknowledges the Natural Science Foundation of Jiangsu Province in China under Grants No. BK20150734, and the Project funded by China Postdoctoral Science Foundation under Grants No. 2014M560419 and No. 2015T80544.
\end{acknowledgments}


\begin{thebibliography}{99}
\bibitem{Kane} M. Z. Hasan and C. L. Kane, Rev. Mod. Phys. \textbf{82}, 3045 (2010).
\bibitem{Zhang} X. L.  Qi  and  S. C. Zhang, Rev. Mod. Phys. \textbf{83}, 1057 (2011).
\bibitem{Bernevig} B. A. Bernevig, T. L. Hughes, and S. C. Zhang, Science \textbf{314}, 1757 (2006).
\bibitem{Molenkamp}  M. K\"{o}nig \emph{et al}., Science \textbf{318}, 766 (2007).
\bibitem{Roth} A. Roth \emph{et al}., Science \textbf{325}, 294 (2009).
\bibitem{Kane2} C. L. Kane and E. J. Mele, Phys. Rev. Lett. \textbf{95}, 226801 (2005).
\bibitem{Wu} C. Wu, B. A. Bernevig, and S. C. Zhang, Phys. Rev. Lett. \textbf{96}, 106401 (2006).
\bibitem{Kim} J. Maciejko, E. A. Kim, and X. L. Qi, Phys. Rev. B \textbf{82}, 195409 (2010).
\bibitem{Richter} V. Krueckl and K. Richter, Phys. Rev. Lett. \textbf{107}, 086803 (2011).
\bibitem{Dolcini} F. Dolcini, Phys. Rev. B \textbf{83}, 165304 (2011).
\bibitem{Loss}  K. Sato, D. Loss, and Y. Tserkovnyak, Phys. Rev. Lett. \textbf{105}, 226401 (2010).
\bibitem{Chen}  W. Chen \emph{et al}., Phys. Rev. Lett. \textbf{109}, 036802 (2012).
\bibitem{Brune} C. Br\"{u}ne \emph{et al}., Nature Phys. \textbf{8}, 485 (2012).
\bibitem{Hasan} M. Z. Hasan, S. Y. Xu, and M. Neupane, arXiv:1406.1040v2.
\bibitem{BF1} H. Murakawa \emph{et al.}, Science \textbf{342}, 1490 (2013).
\bibitem{BF2} Y. Taguchi \emph{et al.}, Science \textbf{291}, 2573 (2001).
\bibitem{BF3} Y. B. Zhang \emph{et al.}, Nature \textbf{438}, 201 (2005).
\bibitem{BF4} J. Park \emph{et al.}, Phys. Rev. Lett. \textbf{107}, 126402 (2011).
\bibitem{BF5} J. G. Analytis \emph{et al.}, Nat. Phys. \textbf{6}, 960 (2010).
\bibitem{BF6} B. B\"{u}ttner \emph{et al.}, Nat. Phys. \textbf{7}, 418 (2011).
\bibitem{BF7} B. Sac\'{e}p\'{e}\emph{ et al.}, Nat. Commun. \textbf{2}, 575 (2011).
\bibitem{BF8} J. Xiong \emph{et al.}, Phys. Rev. B \textbf{86}, 045314 (2012).
\bibitem{Ando1} T. Ando, T. Nakanishi, and R. Saito, J. Phys. Soc. Jpn. \textbf{67}, 2857 (1998).
\bibitem{Ando2} H. Suzuura and T. Ando, Phys. Rev. Lett. \textbf{89}, 266603 (2002).
\bibitem{Edge} J. M. Edge \emph{et al.}, Phys. Rev. Lett. \textbf{110}, 246601 (2013).
\bibitem{Henny} M. Henny \emph{et al.}, Science \textbf{284}, 296 (1999).
\bibitem{Bocquillon} E. Bocquillon \emph{et al.}, Science \textbf{339}, 1054 (2013).
\bibitem{Rothe} D. G. Rothe \emph{et al.}, New J. Phys. \textbf{12}, 065012 (2010).
\bibitem{Japaridze} A. Str\"{o}m, H. Johannesson, and G. I. Japaridze, Phys. Rev. Lett. \textbf{104}, 256804 (2010).
\bibitem{Ojanen} J. I. V\"{a}yrynen and T. Ojanen, Phys. Rev. Lett. \textbf{106}, 076803 (2011).
\bibitem{Sakurai} J. J. Sakurai, \emph{Modern Quantum Mechanics}, 2nd ed. (Addison Wesley, Reading, MA, 1993).
\bibitem{Zhou} B. Zhou \emph{et al}., Phys. Rev. Lett. \textbf{101}, 246807 (2008).
\bibitem{note} With a spatially varying Rashba coefficient $\alpha(x)$, the Rashba SOC term in Eq. (\ref{H0}) takes a strict form of $H_R=-\frac{i}{2}\{\alpha(x),\partial_x\}\sigma_y=-i\alpha(x)\partial_x\sigma_y-\frac{i}{2}\alpha'(x)\sigma_y$. The anticommutator ensures that $H_R$ is Hermitian even when $\alpha(x)$ is nonuniform. The last term can be dropped when $\alpha(x)$ is a smooth function of satisfying condition, $(\ln\alpha)'\ll k$, with $k$ being the relevant wave vector. In the analytical calculation, we adopt this  assumption for simplicity. However, the main results do not depend on such an assumption since the invariance of $H_R$ under time reversal $T$ together with the property of $T^2=-1$ forbids the intra-edge backscattering \cite{Kane2}. Therefore, the slowly varying assumption of $\alpha(x)$ is of not importance to our results, which has been confirmed by the numerical results. Moreover, the Rashba SOC can penetrate into other regions away from the PCs as well, which will not lead to any change of our results due to the same reason.
\bibitem{SM} See Supplemental Material, which includes Refs. \cite{S1,S2,S3,S4}.
\bibitem{S1} M. K\"{o}nig \emph{et al.}, J. Phys. Soc. Jpn. \textbf{77}, 031007 (2008).
\bibitem{S2} J. Li \emph{et al.}, Phys. Rev. Lett. \textbf{102}, 136806 (2009).
\bibitem{S3} Y. Xing and Q. F. Sun, Phys. Rev. B \textbf{89}, 085309 (2014).
\bibitem{S4} S. Datta, \emph{Electronic Transport in Mesoscopic Systems} (Cambridge University Press, Cambridge, England, 1995).
\bibitem{Groth} C. W. Groth \emph{et al.}, New J. Phys. \textbf{16}, 063065 (2014).
\bibitem{Liu} C. X. Liu \emph{et al.}, Phys. Rev. Lett. \textbf{100}, 236601 (2008).
\bibitem{Du} I. Knez, R. R. Du, and G. Sullivan, Phys. Rev. Lett. \textbf{107}, 136603 (2011).
\bibitem{Schmidt} T. L. Schmidt, Phys. Rev. Lett. \textbf{107}, 096602 (2011).
\bibitem{Romeo} F. Romeo and R. Citro, Phys. Rev. B \textbf{90}, 155408 (2014).
\bibitem{Das} S. Das and S. Rao, Phys. Rev. Lett. \textbf{106}, 236403 (2011).
\bibitem{Soori} A. Soori, S. Das, and S. Rao, Phys. Rev. B \textbf{86}, 125312 (2012).
\bibitem{Hou} C. Y. Hou, E. A. Kim, and C. Chamon, Phys. Rev. Lett. \textbf{102}, 076602 (2009).
\bibitem{Adroguer} P. Adroguer \emph{et al.}, Phys. Rev. B \textbf{82}, 081303 (2010).
\end{thebibliography}
\end{document}